\documentclass[11pt]{article}
\usepackage{mathrsfs}
\usepackage{amsmath}
\usepackage{amsfonts}
\usepackage{amssymb, amsmath, cite}
\usepackage{color}

\newtheorem{theorem}{Theorem}[section]

\newcommand\be{\begin{equation}}
\newcommand\ee{\end{equation}}
\newcommand\ber{\begin{eqnarray}}
\newcommand\eer{\end{eqnarray}}
\newcommand\berr{\begin{eqnarray*}}
\newcommand\eerr{\end{eqnarray*}}
\newcommand\Om{\Omega}\newcommand\lm{\lambda}
\newcommand\vp{\varphi}\newcommand\lb{\label}\newcommand\eq{\eqref}
\newcommand\vep{\varepsilon}
\newcommand{\dd}{\mathrm{d}}\newcommand\bfR{\mathbb{R}}\newcommand\bfZ{\mathbb{Z}}
\newcommand\e{\mathrm{e}}\newcommand\pa{\partial}
{\setlength\arraycolsep{2pt}
\newcommand\nn{\nonumber}
\newcommand\E{{\bf E}}\newcommand\D{{\bf D}}\newcommand\B{{\bf B}}\newcommand\HH{{\bf H}}

\newcommand\bea{\begin{eqnarray}}
\newcommand\eea{\end{eqnarray}}

\newcommand{\ii}{\mathrm{i}}

\usepackage{graphicx}
\setlength{\baselineskip}{18pt}

\begin{document}

\title{Nonlinear Problems Inspired by \\ the  Born--Infeld Theory of Electrodynamics}

\author{Yisong Yang\,\footnote{Email address: yisongyang@nyu.edu}\\Courant Institute of Mathematical Sciences\\New York University}

\date{\em In honor of Professor Joel Spruck on the occasion of his retirement\\with affection and admiration}

\maketitle

\begin{abstract}

It is shown that nonlinear electrodynamics of the Born--Infeld theory type may be exploited to shed insight into a few fundamental problems in theoretical physics, including rendering
electromagnetic asymmetry to energetically exclude magnetic monopoles, achieving finite electromagnetic energy to relegate curvature singularities of charged black holes, and providing
theoretical interpretation of equations of state of cosmic fluids via k-essence cosmology. Also discussed are some nonlinear differential equation problems.

\medskip


\end{abstract}

\section{Introduction}
\setcounter{equation}{0}

Fundamental physics thrives and even relies on nonlinearities which often lead to highly challenging nonlinear differential equation problems. For example, the free motion of a particle of mass
$m$ and velocity $v$ is governed in Newtonian mechanics by the Lagrangian action function $\frac12 mv^2$ but that in Einstein's special relativity is by
$mc^2(1-\sqrt{1-v^2/c^2})$ where $c$ is the speed of light in vacuum; in order to obtain a full description of the quantum-mechanical motion of a charged particle, the conventional
partial derivative $\pa_\mu$ with respective to the Minkowski spacetime coordinate $x_\mu$ in the Schr\"{o}dinger equation needs to be replaced by the
gauge-covariant derivative $D_\mu=\pa_\mu -\ii A_\mu$, where $A_\mu$ is a real-valued gauge field with induced electromagnetic field $F_{\mu\nu}=\pa_\mu A_\nu-\pa_\nu A_\mu$, giving
rise to nonlinear interaction between the wave function and gauge field in the coupled theory; 
the Yang--Mills gauge field theory describing weak and strong interactions between subatomic particles is formulated in
terms of matrix-valued field tensors of the form $F_{\mu\nu}=\pa_\mu A_\nu-\pa_\nu A_\mu +[A_\mu, A_\nu]$, where $A_\mu$ is a matrix-valued gauge field  and $[\cdot,\cdot]$ the matrix commutator, introducing nonlinear self-interaction of the gauge field; the gravitational theory of Einstein is built over
a pseudo-Riemannian or Lorentzian manifold that relates the spacetime metric tensor to the matter presence through coupling its Ricci tensor and curvature scalar to the matter stress tensor, which inevitably
gives rise to a highly nonlinear partial differential equation problem. Interestingly, in many situations, even when the original theoretical setups are linear and successful, it often becomes necessary 
to go beyond linear structures and stride into nonlinear realms, both for mathematical and physical reasons. For example, mathematically, although the linear Schr\"{o}dinger equation enables a correct
description of
 the full spectral series of hydrogen or any single-electron nucleus systems, it is difficult to achieve similar levels of understanding for the problem involving more than one electron
in the system, such as
helium, lithium, and other atoms, since the linear Schr\"{o}dinger equation now becomes non-separable. In order to tackle such difficulty typically encountered in
quantum many-body problems, various effective {\em nonlinear} methods have been developed, mainly aimed at understanding ground states. These include the Hartree--Fock method \cite{Cramer}, the Thomas--Fermi
formalism \cite{March,Parr}, and the density functional theory \cite{March,Parr}. Furthermore, physically, although the linear London equations are successful in predicting the Meissner effect,
a signature phenomenon in superconductivity, a full phenomenological description of the physics of superconductivity such as phase transition
versus temperature and applied field and onset of mixed states utilizes a mechanism called
spontaneous symmetry breaking, which generically calls upon a quartic potential density function, resulting in
the Ginzburg--Landau theory of superconductivity \cite{GL}. 

This article surveys and elaborates on a few insights \cite{Yang1,Yang2,Yang3} obtained from a nonlinear theory of electrodynamics, known as the Born--Infeld theory \cite{B1,B2,B3,B4}, extending the
classical linear theory of
electromagnetism of Maxwell. Originally, the Born--Infeld theory was formulated with an aim to overcome the energy divergence problem associated with a Coulomb electric point charge
as a model for electron. In contemporary theoretical physics, this theory and its various generalized forms also arise in the research on superstrings \cite{FT,Ts2,Tsey}  and branes \cite{CM,Gibbons,Ts1}, charged black holes \cite{AG1,AG2,K1,K2,Yang1,Yang3},
and cosmology \cite{Jana,Kam,Nov,Yang1,Yang3}. See \cite{JHOR} for a review on modified gravity theories inspired by the Born--Infeld formalism.  In what follows, we first recall the classical Born--Infeld theory
and its generalization (Section 2) and then present three new developments \cite{Yang1} based on the generalized Born--Infeld theory. These include a generic exclusion of monopoles
in view of the Stone--Weierstrass density theorem (Section 3), 
relegation or regularization of curvature singularities of charged black holes (Section 4), and k-essence realization of equations of state for cosmic fluids (Section 5). 
Detailed explanations of these problems will be given in the beginning paragraphs of the respective sections subsequently.
In Section 6, we summarize the results and consider
some nonlinear differential equation problems of analytic interests inspired by the Born--Infeld theory.

\section{Born--Infeld theory and its generalization}\lb{sec2}
\setcounter{equation}{0}

Consider the 4-dimensional Minkowski spacetime with temporal and spatial coordinates, $x^0=t$ and $(x^i)={\bf x}$, equipped with the metric $(\eta_{\mu\nu})=\mbox{diag}(1,-1,-1,-1)$,
which is used to raise and lower coordinate indices as usual. Then the electromagnetic field $F_{\mu\nu}$ induced from a real-valued gauge field $A_\mu$ may be represented in terms of
the underlying electric field ${\bf E}=(E^i)=(F^{i0})$ and magnetic field ${\bf B}=(B^i)=(-\frac12 \vep^{ijk}F^{jk})$. With this preparation, the Lagrangian action density of the Maxwell
theory of electrodynamics reads
\be\lb{2.1}
s={\cal L}_{\mbox{\tiny M}}=-\frac14 F_{\mu\nu}F^{\mu\nu}=\frac12 ({\bf E}^2-{\bf B}^2).
\ee
On the other hand, recall that, the Lagrangian function of Newtonian mechanics for the motion of a free massive particle is ${\cal L}_{\mbox{\tiny N}}=\frac12 mv^2$, and that the
Lagrangian function of special relativity of Einstein for the particle is
\be\lb{2.2}
{\cal L}_{\mbox{\tiny E}}=mc^2\left(1-\sqrt{1-\frac{v^2}{c^2}}\right)=mc^2\left(1-\sqrt{1-\frac2{mc^2}{\cal L}_{\mbox{\tiny N}}}\right).
\ee
In view of this connection and \eq{2.1}, Born and Infeld \cite{B1,B2,B3,B4} proposed their celebrated Lagrangian free action density to be
\be\lb{2.3}
{\cal L}_{\mbox{\tiny BI}}=b^2\left(1-\sqrt{1-\frac2{b^2}{\cal L}_{\mbox{\tiny M}}}\right)=\frac1\beta\left(1-\sqrt{1-2{\beta}s}\right),
\ee
where $b>0$ is called the Born parameter and $\beta=1/b^2$. Note that this Lagrangian was originally proposed by Born himself in \cite{B1,B2}
without elaboration on its special relativity origin from \eq{2.2}, which was later carried out in \cite{B3,B4}.
 For the generalized Born--Infeld theory, the Lagrangian free action density is taken to assume the normalized form \cite{Yang1}
\be\lb{2.4}
{\cal L}=f(s),\quad f(0)=0,\quad f'(0)=1,
\ee
whose Euler--Lagrange equations are
\be\lb{2.5}
\pa_\mu P^{\mu\nu}=0,
\ee
where the field tensor
\be\lb{2.6}
 P^{\mu\nu}=f'(s) F^{\mu\nu}
\ee
gives rise to the usual electric displacement field $\bf D$ and magnetic intensity field $\bf H$ through the relations ${\bf D}=(D^i)=(P^{i0})$ and ${\bf H}=(H^i)=
(-\frac12\vep^{ijk}P^{jk})$. Note also that, in terms of the dual of $F^{\mu\nu}$, namely $\tilde{F}^{\mu\nu}=\frac12\vep^{\mu\nu\alpha\beta}F_{\alpha\beta}$, there also
holds the Bianchi identity $\pa_\mu \tilde{F}^{\mu\nu}=0$, which supplements \eq{2.5}. Besides, the relation \eq{2.6} may be rewritten  in the form of the constitutive equations
between ${\bf E}, {\bf B}$ and ${\bf D}, {\bf H}$ as
\be\lb{2.7}
{\bf D}=\vep({\bf E},{\bf B}){\bf E},\quad \B=\mu(\E,\B)\HH,\quad \vep({\bf E},{\bf B})=f'(s),\quad \mu(\E,\B)=\frac1{f'(s)},
\ee
where the  quantities $\vep$ and $\mu$ resemble the usual dielectrics and permeability coefficients such that $\vep\mu=1$ realizes the fact that the speed of light in vacuum is unity.
In view of \eq{2.5} and \eq{2.7}, we arrive at the following equations of motion 
\be\lb{2.8}
\frac{\pa \B}{\pa t}+\nabla\times\E={\bf 0},\quad \nabla\cdot\B=0,\quad -\frac{\pa\D}{\pa t}+\nabla\times\HH={\bf0},\quad\nabla\cdot\D=0,
\ee
which are of the identical form of the source-free Maxwell equations among which the first two equations are the Bianchi identity and the other two equations are given by \eq{2.5}.
Moreover, the energy-momentum tensor of the theory \eq{2.4} may be calculated to be
\be\lb{2.9}
T_{\mu\nu}=-f'(s) F_{\mu\alpha}\eta^{\alpha\beta}F_{\nu\beta}-\eta_{\mu\nu}f(s),
\ee
resulting in the Hamiltonian energy density
\be\lb{2.10}
{\cal H}=T_{00}=f'(s)\E^2-f(s).
\ee
Thus, in the case of the classical Maxwell theory with $f(s)=s$ and the Born--Infeld theory with $f(s)$ given by  \eq{2.3}, we have
\bea
{\cal H}&=&\frac12(\E^2+\B^2),\lb{2.11}\\
{\cal H}&=&\frac1\beta\left(\frac1{\sqrt{1-\beta(\E^2-\B^2)}}-1\right)+\frac{\B^2}{\sqrt{1-\beta(\E^2-\B^2)}},\lb{2.12}
\eea
respectively, forming interesting comparisons. In the electrostatic case of a point charge source, we have $\nabla\cdot\D=4\pi q \delta({\bf x})$, giving rise to the nontrivial radial
component of $\D$:
\be\lb{2.13}
D^r=\frac q{r^2},\quad r=|{\bf x}|>0,\quad q>0.
\ee
Inserting \eq{2.13} into \eq{2.7}, we obtain
\be\lb{2.14}
E^r=\frac q{r^2};\quad E^r=\frac q{\sqrt{\beta q^2 +r^4}},
\ee
for the Maxwell and Born--Infeld cases, respectively. In view of \eq{2.11}, \eq{2.12}, and \eq{2.14}, it is seen how the divergence and convergence of
the energy of a point electric charge in the Maxwell theory and Born--Infeld theory follow in respective cases. The same is analogously true for a magnetostatic point charge given by
$\nabla\cdot \B=4\pi g\delta({\bf x})$ with $g>0$. In other words, with regard to energy divergence and convergence of an electric or magnetic point charge, there is perfect
symmetry between electricity and magnetism in the Maxwell theory and Born--Infeld theory. In particular, { there is no mechanism to exclude a monopole in either theory}.

\section{Generic exclusion of monopoles}\lb{sec3}
\setcounter{equation}{0}

 In the previous section, we have seen that, in both Maxwell and  Born--Infeld theories,
electric and magnetic point charges are given equal footings energetically, and the theories offer no 
mechanism to rule out the occurrence of a magnetic point charge, i.e., a monopole \cite{Curie,Dirac}. Although the notion
of monopoles is conceptually important in field theory physics \cite{GO,Pre,Wein}, such purely magnetically charged point particles have never been
observed as isolated particles, although some forms of their simulations in condensed-matter systems have been produced \cite{Gib}.
Here we show that the flexibility in its nonlinearity of the generalized Born--Infeld theory makes it possible to break the described electromagnetic
symmetry so that a finite-energy electric point charge is maintained but a finite-energy magnetic point charge is excluded \cite{Yang1}.
Specifically, we shall see that such a  breakdown of electric and
magnetic point charge symmetry, referred to as electromagnetic asymmetry, may be regarded as a  generic property of  nonlinear electrodynamics. More precisely, it will be established that, for any nonlinear electrodynamics governed by a polynomial 
function, the theory always accommodates finite-energy electric point charges but excludes magnetic ones, although unlike
what is seen in the classical Born--Infeld model, no
upper bound for electric field may be imposed in the current context. The word ``generic" is used to refer to the fact that the set of 
polynomials is dense in the function space of nonlinear Lagrangian functions in view of the Stone--Weierstrass theorem \cite{Stone,Yosida} such that
any model of nonlinear electrodynamics may be approximated in a suitable sense by a sequence of models governed by polynomials.
As a consequence, one might conclude that monopoles are generically ruled out with regard to the finite-energy condition.

To proceed, let $f(s)$ be any nonlinearity function given in \eq{2.4} over its maximum interval of definition with end points $a<b$ where $a$, $b$ may be finite or infinite.
Let $\{[a_k,b_k]\}$ be a sequence of compact intervals satisfying $a_k\to a, b_k\to b$ as $k\to\infty$. For any $\vep>0$, let $k_0\geq3$ (say) be such that
\be\lb{3.1}
\sum_{k>k_0}2^{-k}<\frac\vep2.
\ee
For the interest of our problem, we should assume $a_k<0<b_k$ for all $k$. 
For so fixed $k_0$, since $f(s)$ is not linear, there is a nontrivial polynomial $q(s)$ of the form
\be\lb{3.2}
q(s)=\sum_{i=0}^n a_i s^i,\quad a_0,\dots,a_n\neq0,
\ee
such that in terms of the usual $C^0$-norm over $I_{k_0}=[a_{k_0},b_{k_0}]$ we have
\be
| f''-q|_{I_{k_0}}\equiv\max\left\{|f''(s)-q(s)|: a_{k_0}\leq s\leq b_{k_0}\right\}<\frac{\vep}{(|a_{k_0}|+b_{k_0})^2},
\ee
resulting in the bound
\be\lb{3.4}
|f-p|_{I_{k_0}}<\vep,\quad p(s)=s+\sum_{i=0}^n \frac{a_i s^{i+2}}{(i+1)(i+2)},\quad a_0,\dots,a_n\neq0,
\ee
by integration. Consequently, we have
\be\lb{3.5}
d(f,p)\equiv\sum_{k\geq 2} 2^{-k}\frac{|f-p|_{I_k}}{1+|f-p|_{I_k}}<\vep\sum_{2\leq k\leq k_0}2^{-k}+\sum_{k>k_0}2^{-k}<\vep,
\ee
in view of \eq{3.1} and \eq{3.4}. Thus, measured by the metric $d$, $p$ is in the $\vep$-neighborhood of $f$.

The form of the polynomial function $p(s)$ in \eq{3.4} leads us to consider the nonlinearity function
\be\lb{3.6}
f(s)=s+\sum_{k=2}^n a_k s^k, \quad a_2,\dots,a_n\in\bfR,\quad a_n\neq0.
\ee
In the electrostatic situation, $s=\E^2/2$. Thus we may insert \eq{3.6} into \eq{2.7} to obtain
\be
\D=\left(1+\sum_{k=2}^n\frac{ ka_k }{2^{k-1}}\E^{2(k-1)}\right)\E.
\ee
With \eq{2.13}, we see that the nontrivial radial component $E^r$ of $\E$ away from the origin is given by
\bea
E^r&=&\frac q{r^2},\quad r\gg1,\lb{3.8}\\
E^r&=&\left(\frac{2^{n-1} q}{n a_n}\right)^{\frac1{2n-1}} r^{-\frac2{2n-1}},\quad r\ll1,\lb{3.9}
\eea
asymptotically. It is clear that \eq{3.8} is simply the usual Coulomb law but \eq{3.9} is less singular than the Coulomb law near the origin since $n\geq2$. In order to examine the energy of
this point electric charge, we get from \eq{2.10} the Hamiltonian density
\be\lb{3.10}
{\cal H}=\left(1+\sum_{k=2}^n\frac{ ka_k }{2^{k-1}}\E^{2(k-1)}\right)\E^2-\left(\frac{\E^2}2+\sum_{k=2}^n \frac{a_k}{2^k}\E^{2k}\right).
\ee
Using \eq{3.8} and \eq{3.9} in \eq{3.10}, we arrive at the sharp estimates
\be\lb{3.11}
{\cal H}=\frac {q^2}{2 r^4},\quad r\gg1;\quad {\cal H}=(2n-1)\left(\left[\frac{q^2}{2n^2}\right]^n\frac1{a_n}\right)^{\frac1{2n-1}} r^{-\frac{4n}{2n-1}},\quad r\ll1.
\ee
These results lead to the finiteness of the electric energy
\be\lb{3.12}
E=4\pi\int_0^\infty {\cal H} r^2\dd r, 
\ee
for {\em any} $n\geq2$ as anticipated. On the other hand, for a magnetic point charge, the nontrivial radial component of $\B$ reads $B^r=g/r^2$. Hence, in view of this,
$s=-\B^2/2$,  and
\eq{2.10} and \eq{3.6}, we have
\be\lb{3.13}
{\cal H}=\frac{g^2}{2r^4}+\sum_{k=2}^n \frac{(-1)^{k+1} a_kg^{2k}}{2^k r^{4k}},
\ee
so that ${\cal H}=\mbox{O}(r^{-4n})$ as $r\to0$ and ${\cal H}=g^2/ 2r^4$ as $r\to\infty$, asymptotically. In particular, we conclude in view of \eq{3.12} that divergence of energy always occurs for a magnetic
point charge for {\em any} $n\geq2$ as in the Maxwell theory (corresponding to $n=1$). In other words, we see that  magnetic monopoles are energetically excluded in any polynomial model defined by \eq{3.6}.

We summary our results of this discussion as follows.

\begin{theorem} {\rm ({\bf Monopole exclusion theorem})}
Let the nonlinearity function of the generalized Born--Infeld electrodynamics be defined over the interval $I$, which contains $0$ as an interior point, and taken from the set of 
twice continuously differentiable functions given by
\be
{\cal F}=\{f(s)\,|\, f(0)=0, f'(0)=1\},
\ee
and equipped with the distance metric 
\be
d(f,g)=\sum_{k=2}^\infty 2^{-k} \frac{|f-g|_{I_k}}{1+|f-g|_{I_k}},
\ee
where $I_k=[a_k,b_k]$ is a strictly monotone increasing sequence of compact intervals such that $I_k\to I$ as $k\to\infty$ and $|\cdot|_{I_k}$ is the usual $C^0$-norm. Let $\cal P$ be
the subset of $\cal F$ consisting of polynomial functions which is  dense in $\cal F$ by the Stone--Weierstrass theorem.  The nonlinear electrodynamics
theory formulated with any $f\in{\cal P}$ permits finite-energy static electric point charge sources but does not permit any finite-energy static magnetic point charge sources. In other words,
the nonlinear electrodynamics theory with any $f\in{\cal P}$ rejects magnetic monopoles energetically.
\end{theorem}

To apply the theorem, for the classical Born--Infeld model \eq{2.3}, we may take $I_k=\left[-k,\frac1{2\beta}\left(1-\frac1k\right)\right]$; for the exponential model \cite{H1,H2}
\be\lb{3.16}
f_{\mbox{\tiny exp}}(s)=\frac1\beta(\e^{\beta s}-1),\quad \beta>0,
\ee
we may use $I_k=[-k,k]$. It is interesting to see that although both models permit electric as well as magnetic point charge sources but they do not permit magnetic point charge
sources when approximated by the dense subset $\cal P$. 
On the other hand, for the $\arcsin$-model \cite{K2,K3}
\be
f_{\arcsin}(s)=\frac1\beta\arcsin(\beta s),\quad \beta>0,
\ee
we may choose $I_k=\left[-\frac1{\beta}\left(1-\frac1k\right),\frac1\beta\left(1-\frac1k\right)\right]$. It is known that this model {\em itself} does not permit magnetic point charge sources but only electric ones \cite{K2,K3,Yang1}.
\medskip

Thus, in sense of function-space density and exclusion of magnetic monopoles as stated in the theorem, it is seen that we have revealed a general electromagnetic asymmetry phenomenon,  which does not occur
in the Maxwell and Born--Infeld theories.

\section{Relegation of curvature singularities of charged black holes}
\setcounter{equation}{0}

In the context of relativistic physics, mass and energy are considered equivalent. However, these quantities exhibit themselves rather differently
in general relativity, as evidenced in particular in the study of {\em charged} black holes. For example, in the charged Reissner--Nordstr\"{o}m black hole solution \cite{Carroll,MTW,Wald}
situation, gravity and electromagnetism are treated in such a way that
gravitational mass is finite but electromagnetic energy is infinite. In fact, this latter issue is associated with the structure of the Maxwell equations
in which a point charge carries an energy which is divergent at the spot where the charge resides, say, at the radial origin, $r=0$, as described in Section \ref{sec2}.  More specifically, using $(t,r,\theta,\phi)$ to denote
the coordinates of a spherically symmetric spacetime, then the Reissner--Nordstr\"{o}m metric assumes the form
\be\lb{4.1}
\dd s^2=\left(1-\frac{2GM}r+\frac{4\pi G Q^2}{r^2}\right)\dd t^2-\left(1-\frac{2GM}r+\frac{4\pi G Q^2}{r^2}\right)^{-1}\dd r^2-r^2\left(\dd\theta^2+\sin^2\theta\,\dd\phi^2\right),
\ee
where $Q>0$ is an effective charge parameter, $M>0$ the gravitational mass, $G$ Newton's gravitational constant, and the speed of light again set to be unity. The metric \eq{4.1} is an example of the more general Schwarzschild black hole metric \cite{Yang3}
\be\lb{4.2}
\dd \tau^2=g_{\mu\nu}\dd x^\mu\dd x^\nu=A(r)\dd t^2-\frac{\dd r^2}{A(r)}-r^2\left(\dd\theta^2+\sin^2\theta\,\dd\phi^2\right),
\ee
subject to the asymptotic flatness condition $A(r)\to1$ as $r\to\infty$. For \eq{4.2}, the Brown--York   quasilocal energy \cite{BY} contained within the local region stretched to
the ``radial coordinate distance" $r>0$ is given by
\be\lb{4.3}
E_{\mbox{\tiny ql}}(r)=\frac rG\left(1-\sqrt{A(r)}\right),
\ee
so that the limit
\be\lb{4.4}
E_{\mbox{\tiny ql}}(\infty)=\lim_{r\to\infty} E_{\mbox{\tiny ql}}(r),
\ee
gives rise to the Arnowitt--Deser--Misner (ADM) energy or mass \cite{ADM1959,ADM,ADM1962,Carroll,MTW,Wald} of the system in the full space. Thus, if $A(r)$ has the asymptotic form
\be\lb{4.5}
A(r)=1-\frac{2GM}r +\frac{4\pi G Q^2}{r^\sigma}, \quad r\gg1,
\ee
resembling \eq{4.1}, where  the exponent $\sigma$ is an undetermined positive parameter, then the formula \eq{4.3} leads to
\be\lb{4.6}
E_{\mbox{\tiny ql}}(r)=M-\frac{2\pi Q^2}{r^{\sigma-1}}+\frac{GM^2}{2r}-\frac{2\pi GQ^2 M}{r^\sigma}+\mbox{O}\left(r^{-(2\sigma-1)}\right),\quad r\gg1.
\ee
It is clear from \eq{4.3} and \eq{4.5} that the positivity of $E_{\mbox{\tiny ql}}(r)$ requires $\sigma\geq1$. Consequently, if charge (either electric or magnetic or both) contributes to the ADM mass, we must have
$\sigma=1$, in view of \eq{4.4} and \eq{4.6}. In particular, we see that the effective charge of a Reissner--Nordstr\"{o}m black hole does not contribute to the ADM mass
because now $\sigma=2$.
On the other hand, recall that the usual Kretschmann invariant 
\cite{Henry,MTW} of the metric \eq{4.2} is given by
\be\lb{4.7}
K=R_{\alpha\beta\gamma\delta}R^{\alpha\beta\gamma\delta}=\frac{(r^2 A'')^2+4(rA')^2+4(A-1)^2}{r^4},
\ee
in terms of the Riemann tensor $R_{\alpha\beta\gamma\delta}$,
which measures the curvature singularity of the center of the black hole.
For the Schwarzschild solution (the metric \eq{4.1} with $Q=0$), the singularity is of the type $K\sim r^{-6}$. 
For the Reissner--Nordstr\"{o}m charged black hole solution (the metric \eq{4.1} with $Q\neq0$), on the other hand,  the singularity is seen to be elevated to the type
$K\sim r^{-8}$, which is due to the divergence of
energy of an electric or magnetic point charge in the linear Maxwell electrodynamics. Thus, it will be interesting to know what
the Born--Infeld theory and its generalizations can offer for the ADM mass, which is a global quantity, and for the curvature singularity, which is a local property,
in view of the fact that these nonlinear theories of electrodynamics permit point charges of finite energies. To get a picture on what results may be expected, recall that the
property \eq{2.4} is aimed at recovering the Maxwell theory asymptotically in the weak-field limit. Therefore, for charged black holes generated in the generalized Born--Infeld electrodynamics,
we should still obtain the Reissner--Nordstr\"{o}m metric in leading-orders for $r\gg1$, which renders $\sigma=2$ in \eq{4.5} so that electromagnetism still does not contribute to the
ADM mass. On the other hand,  the divergence of energy of a point charge in the Maxwell theory occurs at the center of the source, $r=0$, and the nonlinear electrodynamics of
the Born--Infeld type serves the purpose of removing this divergence. As a consequence, it will be seen that the curvature singularity of a charged black hole at $r=0$ will also be
relegated or ameliorated systematically \cite{Yang1,Yang2,Yang3}.
In general, we shall see that, regarding curvature singularities, finite mass and finite
electromagnetic energy present themselves at an equal footing so that we always return to the Schwarzschild singularity $K\sim r^{-6}$ for $r\ll1$. 
In other words, we conclude that such a singularity relegation phenomenon is universally valid in generalized
Born--Infeld theories. Furthermore, we show that there is a critical mass-energy condition under which
the Schwarzschild type curvature singularity may further be relegated or even eliminated in a systematic way.
In particular, the Bardeen black hole \cite{AG1,AG2,BV,Paula} and the Hayward black hole \cite{Frolov,Hay,Kumar} belong to this category of the
regular black hole
solutions of the Einstein equations coupled with the Born--Infeld type nonlinear electrodynamics for which the critical mass-energy
condition is embedded into the specialized forms of the Lagrangian or Hamiltonian densities.
A common feature shared by the Bardeen
and Hayward black holes is that the form of their Lagrangian action density can only accommodate one sector of electromagnetism,
that is, either electric field or magnetic field is allowed to be present to model a point charge, but not both, due to the sign
restriction under the radical root operation, in a sharp contrast to the classical Born--Infeld theory  which accommodates both
electricity and magnetism at an equal footing. Hence it is imperative to find a nonlinear electrodynamics model that accommodates both electric and magnetic point charges and at the same time gives rise to regular charged black hole solutions
as in the Bardeen and Hayward models. Indeed, we will see \cite{Yang1,Yang2} that such a model may be obtained by taking the large $n$ limit in a 
naturally formulated binomial model, which is a special situation of the polynomial model  considered 
in Section \ref{sec3} in the context of the monopole exclusion mechanism. That is, the binomial model does not
allow a finite-energy magnetic point charge but its large $n$ limit, which assumes the form of an exponential model proposed earlier
by Hendi in
\cite{H1,H2} in other contexts, accommodates both electric and magnetic point charges, and is free of sign restriction. These general and specific issues of charged black holes are our focus
 in this section.

To proceed, we now use the gravitational metric tensor $g_{\mu\nu}$ to lower or raise coordinate indices. Then the equation \eq{2.5} is replaced by
its curved-space version
\be\lb{4.8}
\frac1{\sqrt{-\det(g_{\alpha\beta})}}\pa_\mu\left(\sqrt{-\det(g_{\alpha\beta})}P^{\mu\nu}\right)=0,
\ee
where $P^{\mu\nu}$ assumes the same form, \eq{2.6},  with due metric tensor modification. In this context, the energy-momentum tensor \eq{2.9} becomes
\be\lb{4.9}
T_{\mu\nu}=-f'(s) F_{\mu\alpha}g^{\alpha\beta}F_{\nu\beta}-g_{\mu\nu}f(s).
\ee
Hence, in terms of the Ricci tensor $R_{\mu\nu}$ associated with $g_{\mu\nu}$ and the trace of $T_{\mu\nu}$ given by $T=g^{\mu\nu}T_{\mu\nu}$, the Einstein equation becomes
\be\lb{4.10}
R_{\mu\nu}=-8\pi G\left(T_{\mu\nu}-\frac12 g_{\mu\nu}T\right).
\ee

For a centrally charged situation with spherical symmetry, it can be shown that the coupled equations  \eq{4.8}--\eq{4.10} have the explicit general solution
\be\lb{4.11}
A(r)=1-\frac{2GM}r+\frac{8\pi G}r\int_r^\infty {\cal H}(\rho) \rho^2\,\dd\rho,
\ee
where $A(r)$ is the metric factor given in \eq{4.2} and $M$  an integration constant which may be taken to be positive to represent a mass. Here the Hamiltonian density
${\cal H}=T^0_0$ happens to assume the same form as that in the flat-space situation such that the quantity
\be\lb{4.12}
E=\int {\cal H}\sqrt{-\det(g_{\alpha\beta})}\,\dd x=4\pi\int_0^\infty {\cal H}(\rho) \rho^2\,\dd\rho
\ee
is the electromagnetic energy. Consequently, since the generalized electrodynamics of the Born--Infeld type always recovers the Maxwell theory for which
$E^r=q/r^2$ and $B^r=g/r^2$ for $r\gg1$ when both electric and magnetic charges are present (a dyonic point charge source), so \eq{2.1} leads to
\be\lb{4.13}
s=\frac{q^2-g^2}{2r^4},\quad r\gg1.
\ee
Using \eq{2.4} and \eq{4.13}, we arrive at
\be\lb{4.14}
{\cal H}=\frac{q^2+g^2}{2r^4},\quad r\gg1.
\ee
Inserting \eq{4.14} into \eq{4.11}, we obtain the result
\be\lb{4.15}
A(r)=1-\frac{2GM}r+\frac{4\pi G(q^2+g^2)}{r^2},\quad r\gg1.
\ee
Thus, asymptotically near spatial infinity, the metric factor of a charged black hole in the generalized nonlinear electrodynamics of the Born--Infeld type is
of the same form as that of the Reissner--Nordstr\"{o}m black hole generated by the Maxwell electrodynamics as given in \eq{4.1}. Therefore, as described earlier, the ADM mass
of the so-constructed charged black hole is simply the Schwarzschild mass $M$.

We next study the behavior of the metric factor $A(r)$ near the center of the mass and charges of the black hole assuming that a finite electromagnetic energy is already achieved in \eq{4.12}. 
Under this assumption, substituting \eq{4.12} into \eq{4.11}, we have
\bea\lb{4.16}
A(r)&=&1-\frac{2G(M-E)}r-\frac{8\pi G}r\int_0^r {\cal H}(\rho)\,\rho^2\,\dd\rho\nn\\
&=&1-\frac{2G(M-E)}r-\frac{2G E(r)}r,\quad r>0,
\eea
where $E(r)$ is the electromagnetic energy contained in the spatial region extending to any radial distance $r>0$. This expression clearly indicates how the curvature singularity at $r=0$
may be relegated by the local energy $E(r)$. 

The preceding paragraphs enable us to conclude with the following theorem.

\begin{theorem} {\rm ({\bf Curvature singularity relegation theorem})} Although electromagnetism does not contribute to the ADM mass of a charged black hole given by
the generalized electrodynamics of the Born--Infeld type, a finite electromagnetic energy generically gives rise to a relegated curvature singularity as measured by
the Kretschmann invariant. More precisely, let $E(r)$ denote the electromagnetic energy contained in a spatial region around the center of the mass and charges of the black hole
within any radial distance $r>0$ satisfying
\be\lb{4.17}
E(r)=4\pi \int_0^r {\cal H}(\rho)\rho^2\,\dd\rho=E_0 r^\kappa,\quad r\ll1,
\ee
in leading order, where $E_0>0$ is a constant and $\kappa$ is a parameter for which the condition $\kappa\geq0$ is assumed to observe the finite-energy condition. For any $\kappa\geq0$ the curvature
singularity is relegated to that of the Schwarzschild singularity, $K\sim r^{-6}$. Besides, under the critical mass-energy condition
\be\lb{4.18}
M=E,
\ee
we have
\be\lb{4.19}
K=K_0 r^{2\kappa -6},\quad r\ll1,
\ee
where $K_0>0$ is a constant. Hence in this latter situation the curvature singularity at $r=0$ is removed when $\kappa\geq3$.
\end{theorem}

The theorem is a consequence of using \eq{4.16} and \eq{4.17} directly in \eq{4.7}.

It will be enlightening to discuss a few examples.

First, we consider the electric point charge source situation in the classical Born--Infeld theory \eq{2.3} for which $\cal H$ is given by \eq{2.12} and \eq{2.14} (second expression) with 
zero $\bf B$. Then \eq{4.17} reads
\be
E(r)=\int_0^r\frac{4\pi q^2}{\rho^2+\sqrt{\rho^4+\beta q^2}}\,\dd\rho=\frac{4\pi q r}{\sqrt{\beta}},\quad r\ll1. 
\ee
This gives $E_0=4\pi q/\sqrt{\beta}$ and $\kappa=1$ in \eq{4.17} so that the curvature singularity is relegated to $K\sim r^{-4}$ for $r\ll1$ under the condition \eq{4.18}.  Moreover, we also have
\be
E=E(\infty)=\frac{4\pi q^{\frac32}}{\beta^{\frac14}}\int_0^\infty\frac{\dd x}{x^2+\sqrt{x^4+1}}=\frac{4\pi^{\frac52}q^{\frac32}}{3\left(\Gamma\left(\frac34\right)\right)^2\beta^{\frac14}},
\ee
which determines the critical mass through \eq{4.18} to achieve the stated relegated curvature singularity.

A magnetically charged black hole in the classical Born--Infeld theory \eq{2.3} enjoys the same properties as an electrically charged one due to
the electromagnetic symmetry of the system.

Next, we consider \cite{Yang1,Yang2} the exponential model \eq{3.16} assuming the critical condition \eq{4.18}.

In the electric point charge situation,  \eq{2.7} relates the nontrivial radial components of $\E$ and $\D$ by the equation $f'_{\mbox{\tiny exp}}(s) E^r=D^r$, which gives us
the implicit relation
\be\lb{4.22}
\e^W W=\delta,\quad W=\beta (E^r)^2,\quad \delta =\beta(D^r)^2.
\ee
Recall that the equation $\e^W W=x$ defines the classical Lambert $W$ function \cite{CG,Hayes} which is analytic for $x>-\e^{-1}$ and
\bea
W(x)&=&\sum_{n=1}^\infty\frac{(-n)^{n-1}}{n!} x^n,\quad \mbox{about } x=0,\lb{4.23}\\
W(x)&=&\ln x-\ln\ln x+\frac{\ln\ln x}{\ln x}+\cdots,\quad x>3.\lb{4.24}
\eea
In view of \eq{4.22}, we have $s=W/2\beta$ such that
\be\lb{4.25}
{\cal H}=f'_{\mbox{\tiny exp}}(s) (E^r)^2-f_{\mbox{\tiny exp}}(s)=\frac1\beta\left(\e^{\frac W2}(W-1)+1\right),\quad W=W(\beta [D^r]^2).
\ee
Inserting $D^r=q/r^2$ into \eq{4.25} and using \eq{4.24}, we have
\be\lb{4.26}
{\cal H}=\frac1\beta\left(\frac{\beta^{\frac12} q}{r^2}\left[\ln\frac{\beta q^2}{r^4}-1\right]+1\right),\quad r\ll1,
\ee
resulting in
\be\lb{4.27}
E(r)=\frac{4\pi qr}{\beta^{\frac12}}\left(-4\ln r+3+\ln\beta q^2\right)+\mbox{O}(r^3),\quad r\ll1.
\ee
In view of \eq{4.16}, \eq{4.27}, and \eq{4.7}, we obtain the sharp curvature estimate
\be
K=\frac{2^{10}\pi^2 G^2 q^2\ln^2 r}{\beta r^4}+\mbox{O} (r^{-4}),\quad r\ll1,
\ee
which significantly relegates the Schwarzschild type curvature singularity as well as that in the noncritical situation, $M\neq E$, of course.

In the magnetic point charge situation, we have $s=-(B^r)^2/2=-g^2/2r^4$ and
\be
{\cal H}=-f_{\mbox{\tiny exp}}(s)=\frac1{\beta}\left(1-\e^{-\frac{\beta g^2}{2r^4}}\right),
\ee
such that 
\be
E(r)=\frac{4\pi}\beta\int_0^r\left(1-\e^{-\frac{\beta g^2}{2\rho^4}}\right)\rho^2\,\dd\rho=\frac{4\pi r^3}{3\beta}-\frac{2^{\frac54}\pi g^{\frac32}}{\beta^{\frac14}}
F\left(\frac{2^{\frac14}r}{\beta^{\frac14}g^{\frac12}}\right),\quad r>0,
\ee
where the smooth function
\be
F(r)=\int_0^r \e^{-\frac1{\rho^4}}\rho^2\,\dd\rho
\ee
vanishes at $r=0$ infinitely fast. That is, for any $m\geq1$, we have $F(r)=\mbox{o}(r^m)$ for $r\ll1$. Consequently, \eq{4.16} gives us
\be
A(r)=1-\frac{8\pi G r^2}{3\beta}+\frac{2^{\frac94}\pi G g^{\frac32}}{\beta^{\frac14}r}F\left(\frac{2^{\frac14}r}{\beta^{\frac14}g^{\frac12}}\right) \equiv 1-\frac{8\pi G r^2}{3\beta}+h(r),
\ee
such that \eq{4.7} renders the result
\be
K=24 a^2+\frac{4(h-2ar^2 )h}{r^4}+\frac{4(h'-4ar)h'}{r^2}+(h''-4a)h'',\quad a=\frac{8\pi G}{3\beta}, \quad r>0.
\ee
Therefore, we have 
\be
K=\frac{2^9\pi^2 G^2}{3\beta^2}+\omega(r),
\ee
where $\omega(r)$ is a smooth function which vanishes at $r=0$ infinitely fast. In particular, the mass and charge center is a regular point of the curvature.

It is interesting to note that in this model there is an electromagnetic asymmetry exhibited through the curvature singularities associated with electrically and magnetically charged
black holes in that the singularity associated with electricity is ameliorated but that with magnetism completely regularized. This phenomenon is in sharp contrast with the 
electromagnetic symmetry observed in the classical Born--Infeld theory \eq{2.3}, on the other hand.

Naturally, curvature singularity may also be deteriorated beyond that of the Reissner--Nordstr\"{o}m black hole. For example, for the polynomial model \eq{3.6}, the Hamiltonian
density of a magnetic point charge is given by \eq{3.13}. Hence from \eq{4.11} we have
\be\lb{4.35}
A(r)=1-\frac{2GM}r+\frac{8\pi G (-1)^{n+1} a_n g^{2n}}{(4n-3)2^n r^{2(2n-1)}},\quad r\ll1,
\ee
in leading orders. From \eq{4.35}, we see that \eq{4.7} gives us $K=C_0/r^{8n}$ for $r\ll1$, which indicates that the curvature singularity of a magnetically charged black hole can be
elevated to an arbitrarily high order in the polynomial model as a consequence of the energy divergence associated with a magnetic point charge source presented in Section \ref{sec3}.
This discussion leads us to the interesting observation that, although the binomial model
\be\lb{4.36}
p_n(s)=\frac1\beta\left(\left[1+\frac{\beta s}n\right]^n-1\right),\quad \beta>0,
\ee
for all $n=1,2,\dots$, gives rise to magnetically charged black holes with increasingly high curvature singularities as $n$ goes up, its limit as $n\to\infty$ on the other hand, 
which is the exponential model \eq{3.16}, brings forth magnetically charged black holes free of curvature singularity, rather surprisingly.

\section{k-essence realization of equations of state for cosmic fluids}
\setcounter{equation}{0}

In cosmology, in order to obtain a theoretical interpretation of the observed accelerated expansion of the universe usually attributed to
the existence of dark energy, a real scalar-wave field,  referred to as  quintessence,  may be introduced to 
give rise to  a hidden propelling force \cite{CDS,Carroll,Dv,RP,Tsu}.
In such a context,  quintessence is governed canonically by the Klein--Gordon model such that the kinetic energy density term is minimal and the potential energy density may be adjusted to render the desired dynamic evolution of the
universe.
When the kinetic density term is taken to be an adjustable nonlinear function of the canonical kinetic energy density,
the scalar-wave matter is referred
to as { k-essence} \cite{CGQ,DT,JMW,PL}. This nonlinear kinetic dynamics is of the form of the
 Born--Infeld type theory and may conveniently be used to provide a field-theoretical interpretation of a
 given equation of state relating the pressure and density of a cosmological fluid. Over such a meeting ground,
we may examine the cosmic fluid contents or interpretations of the Born--Infeld type models
in general settings. 

For generality, we consider a homogeneous and isotropic universe formulated over an $(n+1)$-dimensional spacetime with the line element
\be\lb{5.1}
\dd\tau^2=g_{\mu\nu}\dd x^\mu\dd x^\nu=\dd t^2-a^2(t)\delta_{ij}\dd x^i\dd x^j,
\ee
where $a(t)>0$ is the scale factor resembling the radius of the universe.  The nontrivial components of the Ricci tensor of \eq{5.1} are
\be\lb{5.2}
R_{00}=\frac{n \ddot{a}}a,\quad R_{11}=\cdots=R_{nn}=-a\ddot{a}-(n-1)\dot{a}^2,\quad \dot{a}\equiv\frac{\dd a}{\dd t}.
\ee
We are interested in the evolution of the universe propelled by k-essence realized by a real-valued scalar-matter wave function $\vp$ governed by the Lagrangian action density
\be\lb{5.3}
{\cal L}=f(X)-V(\vp),\quad X=\frac12 g^{\mu\nu}\pa_\mu\vp\pa_\nu\vp.
\ee
If we stay within the domain of an extension of the Klein--Gordon model, we should impose the condition $f(0)=0, f'(0)=1$. However, we shall not restrict ourselves to this
condition in our discussion in general since there is no compelling incentive to preserve the Klein--Gordon model in the weak field limit, $X\sim0$. The Euler--Lagrange equation of \eq{5.3} reads
\be\lb{5.4}
\frac1{\sqrt{|\det(g_{\alpha\beta})|}}\pa_\mu\left(\sqrt{|\det(g_{\alpha\beta})|} f'(X)\pa^\mu\vp\right)+V'(\vp)=0,
\ee
which is to be coupled with the $(n+1)$-dimensional Einstein equation 
\be\lb{5.5}
R_{\mu\nu}=-8\pi G_n\left(T_{\mu\nu}-\frac{g_{\mu\nu}T}{n-1}\right),
\ee
extending \eq{4.10}, where $G_n$ is the Newton gravitational constant over an $n$-space, $R$ the scalar curvature,  $T_{\mu\nu}$ the energy-momentum tensor associated with \eq{5.3} given by 
\be\lb{5.6}
T_{\mu\nu}=f'(X)\pa_\mu\vp\pa_\nu\vp -g_{\mu\nu} (f(X)-V(\vp)),
\ee
and $T=g^{\mu\nu}T_{\mu\nu}$ the trace of the energy-momentum tensor.
For consistency, we assume that $\vp$ is also homogeneous, i.e., spatially independent. Then in view of \eq{5.1} the equation of motion \eq{5.4} becomes
\be\lb{5.7}
\left(a^n f'(X)\dot{\vp}\right){\dot{}} =-a^n V'(\vp),\quad X=\frac12\dot{\vp}^2,
\ee
and the nontrivial components of  \eq{5.6} are
\be\lb{5.8}
T_{00}=\dot{\vp}^2 f'(X)-(f(X)-V(\vp)),\quad T_{11}=\cdots=T_{nn}=a^2(f(X)-V(\vp)),
\ee
so that
\be\lb{5.9}
T=\dot{\vp}^2 f'(X)-(n+1)(f(X)-V(\vp)).
\ee
In view of \eq{5.2}, \eq{5.8}, and \eq{5.9}, we see that \eq{5.5} becomes
\bea
&&\frac{\ddot{a}}a=-\frac{8\pi G_n}n\left(\left[\frac{n-2}{n-1}\right] \dot{\vp}^2 f'(X)+\frac2{n-1}(f(X)-V(\vp))\right),\lb{5.10}\\
&&\frac{\ddot{a}}a+(n-1)\left(\frac{\dot{a}}a\right)^2=\frac{8\pi G_n}{n-1}\left(\dot{\vp}^2 f'(X)-2(f(X)-V(\vp))\right).\lb{5.11}
\eea
On the other hand, representing $T_{\mu\nu}$ in terms of its cosmic fluid description characterized by the fluid density $\rho$ and pressure $P$ following the expression
\be\lb{5.12}
(T_{\mu\nu})=\mbox{diag}(\rho, a^2 P,\dots, a^2 P),
\ee
we have by \eq{5.8} the realizations
\be\lb{5.13}
\rho=\dot{\vp}^2f'(X)-(f(X)-V(\vp)),\quad P=f(X)-V(\vp).
\ee
Furthermore, in the current context, the conservation law $\nabla_\nu T^{\mu\nu}=0$ is reduced into
\be\lb{5.14}
\dot{\rho}+n(\rho+P)\frac{\dot{a}}a=0.
\ee
Using \eq{5.13}--\eq{5.14}, we obtain from \eq{5.10}--\eq{5.11} the Friedmann equation
\be\lb{5.15}
\left(\frac{\dot{a}}a\right)^2=\frac{16\pi G_n}{n(n-1)}\rho=\frac{16\pi G_n}{n(n-1)}\left(\dot{\vp}^2 f'(X)-(f(X)-V(\vp))\right).
\ee
The evolution of the homogeneous and isotropic universe is now governed by the closed system of nonlinear differential equations comprised of \eq{5.7} and \eq{5.15}, which is rather complicated in general.
Nevertheless, some simple examples of the system already offer us great insight into the problem.

For example, the equation \eq{5.7} allows two kinds of constant solutions: (i) $\vp=\vp_0$ and $\vp_0$ is a critical point of $V$ such that $V_0=V(\vp_0)>0$, and (ii)
$\vp=\vp_0=$constant and $V=V_0=$a positive constant. In either situation, we have $\rho=V_0$,  $P=-V_0$, $w=P/\rho=-1$, and \eq{5.15} renders us the solution
\be
a(t)=a(0)\e^{E_0t},\quad E_0=\sqrt{\frac{16\pi G_n V_0 }{n(n-1)}},\quad a(0)>0,\quad t>0,
\ee
which fails to yield a Big Bang cosmology but gives rise to a ``dark energy", $E_0$, responsible for the exponential expansion pattern of the universe. In this maner, $V_0$ is 
identified with
 the cosmological constant $\Lambda$ through $\Lambda=8\pi G_n V_0$. Generalizing this idea, we may introduce a field-dependent cosmological ``constant" by setting
\be\lb{5.17}
\Lambda=\Lambda(\vp)=8\pi G_n V(\vp).
\ee
 With this and noting that, in the presence of cosmological constant $\Lambda$,
 the matter density $\rho_m$ and pressure $P_m$ are related to the effective cosmic fluid density $\rho$ and pressure $P$ through
$\rho=\rho_m+\Lambda/8\pi G_n$ and $P=P_m-\Lambda/8\pi G_n$, respectively, we obtain from \eq{5.13} and \eq{5.17} the realizations
\be\lb{5.18}
\rho_m=\dot{\vp}^2f'(X)-f(X)=2Xf'(X)-f(X),\quad P_m=f(X).
\ee
Consequently, we arrive at the constitutive equation
\be\lb{5.19}
w_m=\frac{P_m}{\rho_m}=\frac{f(X)}{2Xf'(X)-f(X)}.
\ee
In particular, in the Klein--Gordon model situation or quintessence cosmology, we have $f(X)=X$, so that $w_m=1$. In other words, the model describes stiff matter. 
 However, in the context of
k-essence cosmology, the kinetic nonlinearity function $f(X)$ enjoys broad freedom for choice so that it may be used to model some desired evolution patterns of the universe.

For example, for a radiation-dominated universe, we have \cite{Al,Cha} $w_m=1/n$, which is implied by the trace-zero condition, $T=g^{\mu\nu}T_{\mu\nu}=0$,  imposed on the electromagnetic energy-momentum
tensor through its perfect fluid realization \eq{5.12}, such that \eq{5.19} gives rise to the differential equation
\be\lb{5.20}
2X f'(X)=(n+1)f(X),
\ee
which leads to the solution
\be\lb{5.21}
f(X)=X^{\frac{n+1}2}.
\ee
In the physical dimension, $n=3$, this gives us the quadratic model $f(X)=X^2$.

In a general formalism,  \eq{5.19} determines how the quantities $P_m$ and $\rho_m$ are related, referred to as the equation of state of the fluid matter.  Since $P_m$ and $\rho_m$ are
parametrized in terms of the quantity $X$, so is $w_m$. In other words, the equation of state is defined by setting $w_m=w_m(X)$ in terms of the parameter $X$, so that \eq{5.19} renders us
\cite{Yang1}
\be\lb{5.22}
2X\frac{\dd f}{\dd X}=\left(1+\frac1{w_m(X)}\right)f,
\ee
which is a separable equation and may readily be integrated to yield a k-essence model \eq{5.3} to realized the given equation of state. This discussion leads to the following.

\begin{theorem}\lb{th5.1} {\rm ({\bf Realization of equation of state of cosmic fluid by k-essence})} In principle, any equation of state of a cosmic matter fluid may be realized
by a k-essence wave-matter model with a suitable choice of the nonlinearity function of the model determined by integrating a first-order separable differential equation.
\end{theorem}

The importance of this construction is that, in the course of cosmological evolution, it is realistic to have a variable $w_m$ for the equation of state rather than a constant one so that various stages of
the expanding universe may be described by cosmic fluids of suitable, respective, physical characteristics, and that $w_m$ plays the role of an interpolation parameter.

We examine a few examples as illustrations, assuming a constant potential density for simplicity.

Consider the following fractional-powered model \cite{Yang1} containing the classical Born--Infeld theory with $p=1/2$,
\be
f(X)=\frac1{\beta}\left(1-\left[1-\frac{\beta X}p\right]^p\right),\quad 0<p<1,\quad\beta>0,
\ee
for generality,
which enables a Big-Bang scenario with $\rho_m\to \infty$, $P_m\to1/\beta$, as $t\to0$, corresponding to $X\to p/\beta$, and $\rho_m\to 0$, $P_m\to 0$, as $t\to\infty$,
corresponding to $X\to 0$. Since now
\be
w_m(X)=\frac{\left(\frac p\beta-X\right)\left(1-\left[1-\frac{\beta X}p\right]^p\right)}{\left(\frac p\beta+[2p-1]X\right)\left(1-\frac{\beta X}p\right)^p+X-\frac p\beta},
\ee
we have
\be
\lim_{X\to\frac p\beta} w_m(X)=0,\quad \lim_{X\to0} w_m(X)=1.
\ee
Thus the fractional-powered k-essence model gives rise to a universe evolving from a dust-dominated fluid to a stiff-matter-dominated fluid.

In the context of the exponential model \eq{3.16}, the k-essence scalar-matter wave is governed by
\be\lb{5.26}
f(X)=\frac1\beta(\e^{\beta X}-1),\quad\beta>0.
\ee
With \eq{5.26}, we see that \eq{5.19} becomes
\be
w_m(X)=\frac{1-\e^{-\beta X}}{2\beta X-1+\e^{-\beta X}}.
\ee
The Big-Bang solution with expansion indicates $\rho_m,P_m\to\infty$ as $t\to0$ and $\rho_m,P_m\to0$ as $t\to\infty$, corresponding to $X\to\infty$ and $X\to0$, respectively, which lead
to the  limits
\be
\lim_{X\to\infty} w_m(X)=0,\quad \lim_{X\to0} w_m(X)=1,
\ee
again interpolating between dust and stiff matter fluids as in the fractional-powered model stituation.

From \eq{3.6}, we may study the polynomial model \cite{Yang3}
\be\lb{5.29}
f(X)=X+\sum_{i=2}^k a_i X^i,\quad a_2,\dots,a_{k-1}\geq0,\quad a_k>0,\quad k\geq2.
\ee
As in the exponential model \eq{5.26}, the Big-Bang growth picture gives rise to the same asymptotic behavior of the matter density and pressure such that $X\to\infty$ as $t\to0$ and
$X\to0$ as $t\to\infty$. With \eq{5.29}, the equation of state \eq{5.19} becomes
\be\lb{5.30}
w_m(X)=\frac{X+\sum_{i=2}^k a_i X^i}{X+\sum_{i=2}^k (2i-1)a_i X^i},
\ee
rendering the results
\be\lb{5.31}
\lim_{X\to\infty} w_m(X)=\frac1{2k-1},\quad \lim_{X\to0} w_m(X)=1,
\ee
independent of the coefficients in \eq{5.29}. To make sense from the first expression in \eq{5.31}, we set
\be
\frac1{2k-1}=\frac1n,\quad k\geq2.
\ee
That is, we relate the expression to a radiation-dominated universe in an $n$-dimensional space. As a consequence, we arrive at the conclusion
\be\lb{5.33}
n=2k-1,\quad k=2,3,\dots.
\ee
In other words, in order that the polynomial k-essence model gives rise to a universe evolving from a radiation-dominated fluid to a stiff-matter one, the space dimensions
may only be odd. The bottom case $(n,k)=(3,2)$ of the condition \eq{5.33} was noted earlier in \cite{Yang3}, which motivates our study here in a general dimension setting.

Recall that, according to modern cosmology, the universe after the Big Bang should go through a radiation-dominated era in the early universe and then evolve into its
matter-dominated phase in later stages. Thus a correct k-essence model should offer such a description for the expansion of the universe. Our study above indicates that
the fractional-powered model, including the classical Born--Infeld theory, and the exponential model both fail to meet this requirement. However, the polynomial models 
successfully serve this purpose when the dimension of the space and degree of the polynomial functions match correctly. We summary this result as follows.

\begin{theorem}  {\rm ({\bf Dimension selection and unique determination of polynomial model})} Consider the polynomial k-essence model consisting
of \eq{5.3} and \eq{5.29} propelling a homogeneous and isotropic universe through its coupling with the Einstein equation under the line element \eq{5.1}
so that $X\to\infty$ as $t\to0$, corresponding to an infinite density-pressure beginning of the universe, or $\rho_m,P_m=\infty$, and that $X\to0$ as $t\to\infty$,
corresponding to a vanishing density-pressure ending of the universe, or $\rho_m=0,P_m=0$. In order to fulfill the required scenario that the universe starts from
a radiation-dominated era in the limit $t=0$ and ends at a stiff-matter-dominated stage in the limit $t=\infty$, the space dimension $n$ must be an odd integer,
$n=2k-1=3,5,\dots$, corresponding to $k=2,3,\dots$. That is, in this situation, the only polynomial model that can be used to achieve such an evolutionary scenario is of degree  $k=(n+1)/2=2,3,\dots$, or quadratic, cubic,\dots, nonlinearities,
corresponding to odd spatial dimensions, $n=3,5,\dots$, respectively.
\end{theorem}

We mention that another criterion for the relevance of a cosmic fluid model is whether the associated adiabatic squared speed of sound, $c^2_s$,  satisfies $c^2_s\in [0,1]$. That is,
whether it stays within the range of the speed of light. For our problem, this is given in view of \eq{5.29} by
\be\lb{5.34}
c_s^2=\frac{\dd P_m}{\dd\rho_m}=\frac{f'(X)}{f'(X)+2Xf''(X)},
\ee
such that it is seen to satisfy the criterion.

To conclude this section, we work out an example to show how to find the corresponding k-essence model to realize the equation of state of a given cosmic fluid model using the method
described in Theorem \ref{th5.1}. For simplicity and interest, we consider the Chaplygin fluid
defined by
equation of state
\be\lb{5.35}
P_m=-\frac {\gamma}{\rho_m}, \quad \gamma>0.
\ee
Directly inserting  \eq{5.35} into \eq{5.18},  we obtain the differential equation
\be
X (f^2)'=f^2-\gamma,
\ee
whose solution reads
$
f(X)=\sqrt{\alpha X+\gamma}
$ and is well known \cite{Yang1}, where $\alpha>0$ is an integration constant. It is clear that the equation of state \eq{5.35} is satisfied with this solution in view of \eq{5.18}.

\section{Conclusion and outlook}
\setcounter{equation}{0}

In this article, we have seen that nonlinear structures inspired by the Born--Infeld theory of electromagnetism may be explored to shed light on
some fundamental issues of field-theoretical physics, including a monopole exclusion mechanism given by electromagnetic asymmetry introduced by polynomial-type nonlinearity,
relegation of curvature singularity of a charged black hole metric as a consequence of achieving finiteness of electromagnetic energy beyond linear theory, 
k-essence interpretation of the equation of state of any prescribed, hypothetical, cosmic fluid, and determination of spacetime dimension in view of a polynomial k-essence model
chosen. The richness of such
nonlinear structures leads to many future directions, both of theoretical and technical interests, to be pursued further. Below we describe a few handily stated nonlinear differential equation
problems in hope to spark further research interests in the Born--Infeld theory related analytic studies.

\subsection{Dyonic matter equations}

First, recall that it was Schwinger \cite{Sch} who extended the work of Dirac \cite{Dirac} on magnetic monopoles and the associated charge quantization formula to the context of
dyons, hypothetical point particles carrying both electric and magnetic charges, based on the Maxwell theory. Thus, it will be interesting to consider dyons in the Born--Infeld theory as well.
Unfortunately, unlike electric and magnetic point charges, it can be shown that the theory \eq{2.3} does not permit a finite-energy dyon \cite{Yang1} as a dually charged point source.
Therefore, the next question is to study continuously distributed dyonic matter in the Born--Infeld theory \eq{2.3} given by the source equations
\be\lb{6.1}
\nabla\cdot{\bf D}=\rho_e({\bf x}),\quad \nabla\cdot{\bf B}= \rho_m({\bf x}),\quad {\bf x}\in\bfR^3,
\ee
where $\rho_e$ and $\rho_m$ are electric and magnetic charge density distribution functions. In the static current-free situation, \eq{2.8} indicates that $\E$ and $\HH$ are conservative. That is,
there are scalar functions $\phi$ and $\psi$ such that $\E=\nabla\phi$ and $\HH=\nabla\psi$. Inserting these into \eq{2.7}, where $f(s)$ is given by \eq{2.3}, and using \eq{6.1}, we have
\bea
\nabla\cdot\left(\nabla\phi\sqrt{\frac{1-\beta|\nabla\psi|^2}{1-\beta|\nabla\phi|^2}}\right)&=&\rho_e,\lb{6.2}\\
\nabla\cdot\left(\nabla\psi\sqrt{\frac{1-\beta|\nabla\phi|^2}{1-\beta|\nabla\psi|^2}}\right)&=&\rho_m,\lb{6.3}
\eea
which are the Euler--Lagrange equations of the action functional
\be
{\cal A}(\phi,\psi)=\int_{\bfR^3}\left(\frac1\beta\left[1-\sqrt{1-\beta|\nabla\phi|^2}\sqrt{1-\beta|\nabla\psi|^2}\right]+\rho_e\phi+\rho_m\psi\right)\,\dd{\bf x}.
\ee
See \cite{Bonh1,Bonh2,Bonh3,Kies1,Kies2} for results on existence, uniqueness, and regularity of the solution to the electric sector of the problem when $\rho_m=0$ and $\psi=0$ over
$\bfR^n$ with $n\geq3$.
In the source-free situation, $\rho_e,\rho_m\equiv0$, it is unknown what the most general entire solutions to \eq{6.2}--\eq{6.3} are, which is a Bernstein or Liouville type problem. A weaker question
in this context is what the most general solutions are under finite-action condition. See \cite{SSY} for some partial results and general formalism.

To construct finite-energy dyonically charged {\em point} sources, we may consider the second Born--Infeld model \cite{B3,B4} based on an invariance principle, which was actually given the item number (2) and expression (2) as well in \cite{B3}, which in the current
generalized context is governed by the Lagrangian action density
\be\lb{6.5}
{\cal L}=f(s), \quad s=\frac12({\bf E}^2-{\bf B}^2)+\frac{\kappa^2}2({\bf E}\cdot{\bf B})^2,\quad \kappa\geq0.
\ee
The constitutive relation between $\D,\HH$ and $\E,\B$ now reads
\be\lb{6.6}
\left(\begin{array}{c}{\bf D}\\{\bf B}\end{array}\right)=\Sigma({\bf E},{\bf B})\left(\begin{array}{c}{\bf E}\\{\bf H}\end{array}\right),
\quad \Sigma({\bf E},{\bf B})\equiv\left(\begin{array}{cc}f'(s)(1+\kappa^4({\bf E}\cdot{\bf B})^2)& \kappa^2({\bf E}\cdot{\bf B})\\ \kappa^2({\bf E}\cdot{\bf B})&\frac1{f'(s)}\end{array}\right),
\ee
such that the matrix $\Sigma({\bf E},{\bf B})$ contains the dielectrics and permeability information of the system and that the property $\det(\Sigma({\bf E},{\bf B}))=1$ resembles the
constraint that the speed of light in vacuum is normalized to unity. This theory enables us to obtain finite-energy dyonically charged point sources, thus restoring electromagnetic symmetry
in the quadratic model,
in particular, and dyonically charged black holes
\cite{Yang2,Yang3} with relegated curvature singularities, as in the first
Born--Infeld theory context, modeled over \eq{2.3} which was given the item number (1) and
expression (1) in \cite{B3}, with either electric or magnetic charges but not both then. In this context, some families of static dyonic matter equations of the form \eq{6.2}--\eq{6.3} are
also derived \cite{Yang3}, which are of analytic challenge.

\subsection{Abelian Higgs model inspired by Born--Infeld theory}

Next, an interesting subject concerns the Abelian Higgs model subject to the Born--Infeld electrodynamics, which is defined by the Lagrangian action density
\be\lb{6.7}
{\cal L}=\frac1\beta\left(1-\sqrt{1-2\beta{\cal L}_{\mbox{\tiny M}}}\right)+\frac12(\overline{D_\mu\phi })(D^\mu\phi)-V(|\phi|^2),
\ee
where $\phi$ is a complex-valued scalar field, $D_\mu\phi=\pa_\mu\phi-\ii A_\mu\phi$ the gauge-covariant derivative, and $V\geq0$ a potential density function. In the two-dimensional static limit
and under the temporal gauge $A_0=0$ (in the classical Abelian Higgs theory, finite-energy condition implies the temporal gauge in two-spatial dimensions, so that the theory must be purely magnetic without electricity. This statement is known as the Julia--Zee theorem \cite{JZ,SY}. In the Born--Infeld theory case, it is of interest to study whether the same
statement would be true),  the Euler--Lagrange equations of
\eq{6.7} are
\bea
D_i D_i\phi&=&2V'(|\phi|^2)\phi,\lb{6.8}\\
\pa_j\left(\frac{F_{ij}}{\sqrt{1+\beta F_{12}^2}}\right) &=&\frac\ii2(\phi\overline{D_i\phi}-\overline{\phi}D_i\phi),\lb{6.9}
\eea
where $i,j=1,2$. The solutions of these equations are also the critical points of the energy functional
\be\lb{6.10}
E(\phi, A)=\int_{\bfR^2}\left(\frac1\beta\left[\sqrt{1+\beta F^2_{12}}-1\right]+\frac12 |D_1\phi|^2+\frac12|D_2\phi|^2+V(|\phi|^2)\right)\dd x,
\ee
where $A=(A_i)$, so that a finite-energy solution of \eq{6.8}--\eq{6.9} satisfies the following Derrick--Pohozaev type identity
\be
\int_{\bfR^2}\left(\frac1\beta\left[\sqrt{1+\beta F^2_{12}}-1\right]+V(|\phi|^2)\right)\dd x=\int_{\bfR^2}\frac{F_{12}^2}{\sqrt{1+\beta F^2_{12}}}\dd x.
\ee
As in the formalism of the Abelian Higgs theory, we assume that there is a spontaneously broken $U(1)$ symmetry realized by the vacuum manifold given by $V=0$ at $|\phi|^2=\phi_0^2>0$ which may be taken to be unity for convenience. That is, $V(1)=0$ and $\phi_0=1$.
Now since $|\phi(x)|\to1$ as $|x|\to\infty$ for a solution of \eq{6.8}--\eq{6.9}, we see that
\be
\Gamma=\frac{\phi}{|\phi|}: \quad S^1_R\to S^1
\ee
is well defined when $R>0$ is large enough, where $S^1_R$ denotes the circle in $\bfR^2$
centered at the origin and of radius $R$. Therefore the map $\Gamma$ may be viewed as an element in
the fundamental group
$
\pi_1(S^1)=\bfZ
$
and represented by an integer $N$. In fact, this integer $N$ is the winding number of $\phi$
around $S^1_R$ and may be expressed by the integral
\be\lb{6.13}
N=\frac1{2\pi\ii}\int_{S^1_R}\dd \ln \phi.
\ee
The continuous dependence of the right-hand side of  \eq{6.13} with respect to $R$ indicates that this quantity is independent of $R$ since the left-hand side of \eq{6.13}
is an integer. Thus we arrive at the following magnetic flux quantization condition
\be\lb{6.14}
\Phi=\int_{\bfR^2} F_{12}\,\dd x=2\pi N,
\ee
as a consequence of the limit
\bea
&&\bigg|\int_{|x|\leq R} F_{12}\, \dd x+\ii\int_{|x|=R} \dd\ln\phi\bigg|=
\bigg|\int_{|x|=R}
A_i\, \dd x_i+\ii\int_{|x|=R}\,\phi^{-1}\pa_i\phi\,\dd x_i\bigg|\nn\\
&&\leq\int_{|x|=R} |\phi^{-1}|(|D_1\phi|+|D_2\phi|)\,\dd s
\to 0,\quad \mbox{as }R\to\infty,
\eea
since $|D_1\phi|,|D_2\phi|\to0$ as $|x|\to\infty$ exponentially fast. The topological number $N$ given in \eq{6.13} or \eq{6.14} is called the vortex number of the solution.
Conversely, we ask  whether for any given $N\in\bfZ$ there is a solution to
the topologically constrained minimization problem
\be\lb{6.16}
E_N\equiv \inf\bigg\{E(\phi,A)\,\bigg|\, \int_{\bfR^2} F_{12}\,\dd x= 2\pi N\bigg\}.
\ee
As in the classical Abelian Higgs theory \cite{JT}, this is a difficult problem, although a self-duality structure may be explored as discovered in \cite{SH} to offer a partial understanding of the
problem. To see how, we use the identity
\be
|D_1\phi|^2+|D_2\phi|^2=|D_1\phi\pm\ii D_2\phi|^2\pm\ii(D_1\phi\overline{D_2\phi}-\overline{D_1\phi}D_2\phi),
\ee
to rewrite the Hamiltonian density of \eq{6.10} as
\bea\lb{6.18}
{\cal H}&=&\frac{\left(F_{12}\pm\frac12\sqrt{1+\beta F_{12}^2}\,(|\phi|^2-1)\right)^2}{2\sqrt{1+\beta F_{12}^2}}+\frac{\left(\sqrt{1+\beta F_{12}^2}\sqrt{1-\frac\beta4(|\phi|^2-1)^2}-1\right)^2}{2\beta\sqrt{1+\beta F_{12}^2}}\nn\\
&&-\frac1\beta\mp \frac12 F_{12}(|\phi|^2-1)+\frac1\beta\sqrt{1-\frac\beta4(|\phi|^2-1)^2}\nn\\
&&+\frac12|D_1\phi\pm\ii D_2\phi|^2\pm\frac\ii2(D_1\phi\overline{D_2\phi}-\overline{D_1\phi}D_2\phi)
+V(|\phi|^2).
\eea
Besides, in view of the commutator or curvature relation $(D_1 D_2-D_2D_1)\phi=-\ii F_{12}\phi$, we see that the current density
\be
J_i=\frac\ii2(\phi\overline{D_i\phi}-\overline{\phi}D_i\phi),\quad i=1,2,
\ee
gives rise to the vorticity field
\be\lb{6.20}
J_{12}=\pa_1 J_2-\pa_2 J_1=\ii (D_1\phi \overline{D_2\phi}-\overline{D_1\phi}D_2\phi)-|\phi|^2 F_{12}.
\ee
We can now {\em choose}
\be\lb{6.21}
V(|\phi|^2)=\frac1\beta\left(1-\sqrt{1-\frac\beta4(|\phi|^2-1)^2}\right),
\ee
under the condition $\beta<4$.
 Then $V(1)=0$ as desired so that the $U(1)$-symmetry is spontaneously broken. Using \eq{6.20}--\eq{6.21} in \eq{6.18}, we have
\be\lb{6.22}
{\cal H}\geq \pm \frac12 (F_{12}+J_{12}).
\ee
 Furthermore, since $D_1\phi$ and $D_2\phi$ vanish at infinity rapidly, we have $\int_{\bfR^2} J_{12}\,\dd x=0$. Hence, with $N=\pm |N|$, we see that \eq{6.22} leads us to
the topological lower bound
\be\lb{6.23}
E(\phi,A)=\int_{\bfR^2}{\cal H}\dd x\geq\pm\frac12 \int_{\bfR^2} F_{12}\,\dd x=\pi|N|,
\ee
by \eq{6.14} and \eq{6.22}, and this lower bound is saturated when the following Bogomol'nyi \cite{Bo} type equations are satisfied:
\bea
&&F_{12}\pm\frac12\sqrt{1+\beta F_{12}^2}\,(|\phi|^2-1)=0,\lb{6.24}\\
&&\sqrt{1+\beta F_{12}^2}\sqrt{1-\frac\beta4(|\phi|^2-1)^2}-1=0,\lb{6.25}\\
&&D_1\phi\pm\ii D_2\phi=0.\lb{6.26}
\eea
Similar equations also appear fruitfully in the Yang--Mills theory \cite{Actor,Har,Hitchin,PS,R,SSY2,SSY3,STY,Witten}.
It is clear that \eq{6.24} implies \eq{6.25} so that these two equations may be compressed into one,
\be\lb{F}
F_{12}=\pm\frac{1-|\phi|^2}{2\sqrt{1-\frac\beta4(|\phi|^2-1)^2}}.
\ee
It can be shown that $\phi$ satisfies the condition $0\leq|\phi|^2\leq 1$, $-1/2\leq F_{12}\leq1/2$, and $F_{12}=\pm1/2$ at $\phi=0$. In other words, the magnetic or vorticity field $F_{12}$
acquires its greatest  strength $\pm1/2$ at the zeros of $\phi$ which represent ``vortex points".
 Moreover, \eq{6.26} implies that $\phi$ is locally holomorphic or anti-holomorphic up to a nonvanishing smooth factor such that the zeros of $\phi$ are
algebraic, that is, the zeros of $\phi$ are isolated, which are $p_1,\dots,p_k$, with respective integer multiplicities, $n_1,\dots,n_k$, summing up to $|N|$, $n_1+\cdots+n_k=|N|$,
so that a charge $N$ configuration indeed gives rise to an $N$-vortex solution. Resolving 
\eq{6.26} away from $p_1,\dots,p_k$, we have $2F_{12}=\mp\Delta|\phi|^2$.  Thus, using $u=\ln|\phi|^2$ and taking account of the zeros $p_1,\dots,p_k$ of $\phi$ and their respective multiplicities
$n_1,\dots,n_k$, we obtain from \eq{F} the equation
\be\lb{6.27}
\Delta u=\frac{\e^u-1}{\sqrt{1-\frac\beta4(\e^u-1)^2}}+4\pi\sum_{l=1}^k n_l \delta_{p_l}(x),\quad x\in\bfR^2,
\ee
subject to the boundary condition $u=0$, corresponding to $|\phi|^2=1$, at infinity.
For this equation, an existence and uniqueness theorem for its solution has been established \cite{Yang0} which gives rise to the unique solution
up to gauge transformations to the optimization problem \eq{6.16} with 
\be\lb{6.28}
E_N=\pi|N|,
\ee
where $V$ assumes the special form \eq{6.21}, realizing prescribed zeros with associated multiplicities as point vortices. Note that, when $\beta=0$, the equation \eq{6.27} reduces into that in the classical Abelian Higgs theory \cite{JT,T}, sometimes referred to
as the Taubes equation \cite{Gud,Man}. In a slightly more general situation where
\be
V(|\phi|^2)=\frac\lm\beta\left(1-\sqrt{1-\frac\beta4(|\phi|^2-1)^2}\right),\quad\lm>0,
\ee
we may rewrite the corresponding energy functional \eq{6.10} as $E^\lm(\phi, A)$ so that the quantity given in \eq{6.16} is denoted by $E^\lm_N$. It is clear that
\be\lb{6.30}
\lm E^1(\phi,A)\leq E^\lm(\phi,A)\leq E^1(\phi,A), \quad \lm\leq1;\quad  E^1(\phi,A)\leq E^\lm(\phi,A)\leq\lm E^1(\phi,A),\quad\lm\geq1. 
\ee
Consequently, in view of \eq{6.28} or $E^1_N= \pi|N|$ and \eq{6.30}, we get the energy estimate
\be\lb{6.31}
\min\{1,\lm\} \pi|N|\leq E^\lm_N\leq\max\{1,\lm\}\pi|N|,
\ee
in terms of the topological charge $N$. In particular, the left-hand side of \eq{6.31} indicates an {\em energy gap}. That is, the interval $\left(0, \min\{1,\lm\} \pi\right)$ does not contain any energy point
of the system.

The productive study of the self-dual system \eq{6.24}--\eq{6.26} suggests that we extend the method of \cite{SH} shown above to obtain self-dual reductions for various
Born--Infeld inspired Abelian Higgs models for which the Born--Infeld electromagnetic action density \eq{2.3} is extended to assume the general form \eq{2.4}. Apparently, the main difficulty here is how to overcome
the limitation associated with the completion-of-squares procedure used in \eq{6.18} in order to make the topological invariant \eq{6.14} stand out. Any new constructions beyond \eq{6.10}
will be interesting. See \cite{LY1} for a construction of gauged harmonic maps along the line of the Born--Infeld theory \eq{2.3} in a similar spirit.

Now return to the model \eq{6.10}. Using polar coordinates $r,\theta$ on $\bfR^2$, a radially symmetric $N$-vortex solution $(\phi,A_i)$ is given by the ansatz
\be\lb{6.32}
\phi(x)=u(r)\e^{\ii N\theta},\quad A_i(x)=Nv(r)\varepsilon_{ij}\frac{x^j}{r^2},\quad i,j=1,2,\quad N\in{\bfZ},
\ee
where $u,v$ are real-valued functions satisfying the regularity condition
$u(0)=v(0)=0$. Moreover, inserting \eq{6.32} into \eq{6.10}, we have
\be\lb{6.33}
E(u,v)=2\pi\int_0^\infty\left(\frac1\beta\left[\sqrt{1+\beta N^2 \frac{(v')^2}{r^2}}-1\right]+\frac{(u')^2}2+\frac{N^2}{2r^2} u^2 (v-1)^2+V(u^2)\right) r\dd r.
\ee
 In view of this and the structure of \eq{6.33}, we arrive at the full set of boundary conditions for $u,v$ as follows:
\be\lb{6.34}
u(0)= v(0)=0,\quad u(\infty)=v(\infty)=1,
\ee
where $N\neq0$.
Subject to \eq{6.34} and varying \eq{6.33}, we get the Euler--Lagrange equations
\bea
u''+\frac{u'}r&=&\frac{N^2}{r^2} (v-1)^2 u+2V'(u^2) u,\lb{6.35}\\
\left(\frac{v'}{r\sqrt{1+\beta N^2 \frac{(v')^2}{r^2}}}\right)' &=& \frac{u^2(v-1)}r,\lb{6.36}
\eea
which are also the radially reduced version of \eq{6.8}--\eq{6.9}. It will be interesting to develop an existence theory for the solutions to \eq{6.35}--\eq{6.36} as critical points of the functional
\eq{6.33} subject to the boundary condition \eq{6.34}. We expect to recover the $N$-vortex solutions of the classical Abelian Higgs model \cite{BC} when $\beta\to 0$. This problem is of independent
interest.

Furthermore, a truly one-dimensional reduction (domain walls) of the problem is worth considering as well. In this setting, we may assume that the fields $\phi$ and $A_i$ depend on $x^1=x$ only,
$\phi=f(x)$ is real-valued, and $A_1=0, A_2=a(x)$. So \eq{6.26} becomes $f'\pm a f=0$. In the nontrivial situation, $f$ never vanishes such that we may assume $f>0$, which gives us
$a=\mp(\ln f)'$. Besides, we have $F_{12}=a'$. In view of these, we get from \eq{F} the self-dual domain-wall equation
\be
u''=\frac{\e^u-1}{\sqrt{1-\frac\beta4(\e^u-1)^2}},\quad u=2\ln f.
\ee
Following \cite{Bol},  boundary conditions of interest describing relevant phase transition phenomena include $u(-\infty)=0, u(\infty)=-\infty$ and $u(\pm\infty)=-\infty$. When $\beta=0$, the
equation is a one-dimensional Liouville type equation \cite{Liouville} and can be integrated \cite{CCY}. Whether the equation may be integrated in the Born--Infeld case, $\beta>0$, is to be
studied. More generally, subject to the same domain-wall ansatz, the equation \eq{6.8}--\eq{6.9} become
\bea
f''-a^2f&=& 2V'(f^2)f,\lb{39}\\
\left(\frac{a'}{\sqrt{1+\beta (a')^2}}\right)'&=&f^2 a,\lb{40}
\eea
and the energy \eq{6.10} assumes the form
\be\lb{41}
E(a,f)=\int_{-\infty}^\infty\left(\frac1\beta\left[\sqrt{1+\beta (a')^2}-1\right]+\frac12(f')^2+\frac12 a^2 f^2+V(f^2)\right)\dd x.
\ee
It is clear that \eq{39}--\eq{40} are the Euler--Lagrange equations of the energy functional \eq{41}. As an extension to the Ginzburg--Landau equations for superconductivity theory,
boundary value problems over a finite interval are of interest too. For example, for the phase transition between the normal state at $x=-1$ and the superconducting state at $x=1$
(say), we may impose $f(-1)=0, a'(-1)=H_0>0$ (the normal magnetic phase, where $H_0$ represents an applied magnetic field) and $f(1)=1, a'(1)=0$ (the superconductive Meissner effect phase), respectively.

\subsection{MEMS equations based on Born--Infeld electromagnetism}

Finally, we consider a nonlinear differential equation problem associated with the Born--Infeld theory inspired formalism
of electrostatic actuation arising in the study of microelectromechanical systems, known as MEMS \cite{PB}.  To proceed, recall that the Coulomb law states that the electrostatic force
$F$ between two charges, $q_1$ and $q_2$, placed at a distance, $r$, apart is given by $F(r)=q_1 q_2/r^2$. Now assume $q_1$ and $q_2$ are uniformly distributed over two parallel tiny plates.
If the two charges are equal in magnitude but of opposite signs, $q_1=-q_2=q$, and brought together to a finite separation distance $r>0$ from infinite separation, then the potential
acquired is the work done given by
$U(r)=\int_r^\infty F(\rho)\,\dd \rho=-q^2/r$ so that  $F(r)=-U'(r)$. To maintain the charges $q$ and $-q$ in the two plates, an electric field is applied and measured in voltage $V$, 
which is proportional to $q$. Thus, with normalization and within small separation oscillation assumption, we may take $q=V$ such that $U$ becomes $U=-V^2/r$. When the plates are subject to deformation extended over a planar region $\Omega$
so that $r$ is described by $r=L+u(x)$ ($x\in\Omega$) where $L>0$ is the distance between the two plates in absence of deformation and $u(x)$ represents the vertical deformation 
amount fluctuating about $u=0$, then $U$ is given by an integral instead:
\be\lb{6.37}
U=-\int_\Omega \frac{V^2}{L+u}\,\dd x,\quad\Omega\subset\bfR^2.
\ee
 On the other hand, in the situation of the Born--Infeld theory, from \eq{2.14}, we know that the force $F$ is modified into 
$
F(r)=-q E^r=-q^2/\sqrt{\beta q^2+r^4}$ such that in the absence of plate deformation, the model gives us
\be\lb{6.38}
U(r)=-\int_r^\infty\frac{ V^2}{\sqrt{\beta  V^2+\rho^4}}\,\dd\rho.
\ee
Hence, when deformation is considered, we have
\be\lb{6.39}
U=\int_r^\infty F(\rho)\dd\rho=-\int_\Om \int_{L+u(x)}^\infty\frac{ V^2}{\sqrt{\beta V^2+\rho^4}}\,\dd\rho\dd x,
\ee
replacing \eq{6.37}. Therefore, adding the stretching, bending, and elastic energies to the electrostatic potential energy \eq{6.39},
 we come up with the total elastic-electrostatic energy functional
\be\lb{6.40}
E(u)=\int_\Om\left(\frac T2|\nabla u|^2+\frac D2|\Delta u|^2+\frac\kappa2 u^2-\int^\infty_{L+u(x)}\frac{V^2}{\sqrt{\beta V^2+\rho^4}}\,\dd\rho\right)\dd x,
\ee
where $T>0$ is the tension constant, $D>0$ relates to the Young modulus, $\kappa>0$ the elastic constant, and $V>0$ is an effective applied voltage. Varying $u$ in \eq{6.40}, we arrive at the following 
Born--Infeld theory modified equation:
\be\lb{6.41}
T\Delta u-D\Delta^2u=\kappa u+\frac{V^2}{\sqrt{\beta V^2+(L+u)^4}},
\ee
governing the static configuration of a MEMS electric actuator, whose $\beta=0$ limit has been studied in \cite{LY2}. Furthermore, to investigate the dynamics of such a system, we may
regard \eq{6.40} as the total potential energy so that the associated Lagrangian action functional is given by
\be\lb{6.42}
I(u)=\int_\Om \frac12 u_t^2\dd x-\int_\Om\left(\frac T2|\nabla u|^2+\frac D2|\Delta u|^2+\frac\kappa2 u^2-\int^\infty_{L+u(x)}\frac{V^2}{\sqrt{\beta V^2+\rho^4}}\,\dd\rho\right)\dd x.
\ee
Varying $u$ in \eq{6.42}, we obtain the equation of motion of the system:
\be\lb{6.43}
u_{tt}=T\Delta u-D\Delta^2 u-\kappa u-\frac{V^2}{\sqrt{\beta V^2+(L+u)^4}}.
\ee
A boundary condition of interest  is the ``pinned" or Navier boundary condition $u=\Delta u=0$ on $\pa\Om$. The simplified homogeneous case
of the problem for which $u$ is spatially independent is also of interest. In this situation, the wave equation \eq{6.43} becomes a nonlinear ordinary differential equation:
\be\lb{6.44}
\ddot{u}+\kappa u+\frac{V^2}{\sqrt{\beta V^2+(L+u)^4}}=0.
\ee
In the Maxwell theory limit, $\beta=0$, it is shown in \cite{YZZ} that, there is an explicitly determined critical voltage $V_c>0$, called the pull-in voltage \cite{LE}, such that below $V_c$ the
equation has a periodic solution oscillating between two ``stationary" states $u(0)=0, \dot{u}(0)=0$ and $u(t_s)=-u_s<0, \dot{u}(t_s)=0$ where $t_s>0$ is called the stagnation time
\cite{LE} which gives rise to the period of the oscillation, $\tau=2t_s$;  above $V_c$ the solution $u$ monotonically goes to its limiting position in finite time; and at $V_c$, the solution $u$ monotonically approaches
its limiting position as $t\to\infty$. In other words, oscillatory vibration of the electric actuator occurs if and only if $V<V_c$. Moreover, similar conclusions may be established when nonlinear elasticity is also considered for the system. Due to the microscopic-scale nature of MEMS
devices, it
will be useful to modify the divergent Maxwell theory formalism with the convergent Born--Infeld theory formalism to describe MEMS electric actuators. It is this consideration that motivates
a systematic study of the associated nonlinear differential equations problems such as \eq{6.41}, \eq{6.43}, and \eq{6.44} along \cite{LY2,YZZ} and the references therein.

\end{document}